\documentclass[11pt]{article}

\usepackage[english]{babel}	
\usepackage[utf8x]{inputenc}	
\usepackage{enumerate}
\usepackage{typearea}
\usepackage{pdfpages}
\usepackage[]{todonotes}
\usepackage{subcaption}
\usepackage{verbatim}
\usepackage{comment}
\usepackage{mathtools}
\usepackage[hidelinks]{hyperref}
\usepackage{ulem}

\usepackage{geometry}
\newgeometry{textwidth=16cm, textheight=24cm}

\usepackage{enumitem}
\usepackage{amsfonts, amsmath, amssymb, amsthm, dsfont}
\usepackage{amsthm}			
\usepackage{thumbpdf}	
\usepackage{natbib}
\allowdisplaybreaks

\NewCommandCopy{\amsthmproof}{\proof}
\RenewDocumentCommand{\proof}{o}{%
  \IfNoValueTF{#1}
    {\amsthmproof}
    {\amsthmproof[\proofname\ #1]}%
}

\usepackage{booktabs} 
\usepackage[noblocks]{authblk}
\usepackage{ulem}

\usepackage{pgf, tikz}
\usepackage{tikz-cd}		
\usepackage{bbm}

\theoremstyle{definition}
\newtheorem{definition}{Definition}[section]
\newtheorem{assumption}{Assumption}
\newtheorem*{assumption*}{Assumption}

\newtheorem*{condition*}{Condition}

\theoremstyle{plain}
\newtheorem{theorem}[definition]{Theorem}
\newtheorem{proposition}[definition]{Proposition}
\newtheorem{lemma}[definition]{Lemma}

\theoremstyle{remark}

\usepackage{graphicx}

\newcommand{\eps}{{\varepsilon}}
\renewcommand{\phi}{\varphi}
\renewcommand{\theta}{\vartheta}

\newcommand{\R}{\mathbb{R}}

\newcommand{\N}{\mathbb{N}}

\newcommand{\pr}{\mathbb{P}}       
\newcommand{\ex}{\mathbb{E}}       
\newcommand{\var}{\textnormal{Var}}

\newcommand{\Dc}{\mathcal{D}}

\newcommand{\Fc}{\mathcal{F}}
\newcommand{\Gc}{\mathcal{G}}
\newcommand{\Nc}{\mathcal{N}}
\newcommand{\Oc}{\mathcal{O}}

\newcommand{\Uc}{\mathcal{U}}

\newcommand{\diff}{{\,\mathrm{d}}}
\newcommand{\convw}{\Rightarrow}

\newcommand{\id}{\mathds{1}}

\title{Self-normalised change point tests that remain pivotal under non-stationarity}
\author[1]{Florian Heinrichs}
\author[2,3]{Fabian Mies}

\affil[1]{FH Aachen -- University of Applied Sciences, Department of Medical Engineering and Technomathematics}
\affil[2]{Delft University of Technology, Delft Institute of Applied Mathematics}
\affil[3]{Ruhr University Bochum, Faculty of Mathematics}

\date{}

\begin{document}

\maketitle

\begin{abstract}
    Self-normalising test statistics for time series circumvent the lag-specification of long-run variance estimators by employing an inconsistent estimator for studentization.
This leads to asymptotically pivotal test distributions under stationarity, but not for nonstationary sequences. 
Focusing on the multiple changepoint problem, we propose an alternative self-normalised test statistic which is asymptotically pivotal even if the noise is nonstationary. 
Central to our argument are statistical functionals which are invariant under temporal reparametriation, i.e.\ clock-changes. 
Simulations show that the proposed procedure is more conservative than competing bootstrap procedures, but considerably more powerful against multiple changes and more reliable in terms of size for highly nonstationary time series in larger samples.

\end{abstract}

\section{Introduction} \label{sec:intro}

In the nonparametric regression model $X_i = \mu(i/n) + \epsilon_i$ for a non-random mean function $\mu$ and centered random variables $\epsilon_i$, we consider the changepoint problem in the form of a test for the null hypothesis
\begin{align*}
    \mathbb{H}_0: \mu \text{ is constant } \quad \text{against}\quad \mathbb{H}_1: \mu \text{ is a step-function with one jump}.
\end{align*}
This testing problem is commonly approached via the CUSUM statistic $T_n=T^{\textsc{CUSUM}}(S_n) = \max_{t\in [0,1]} |S_n(t)-t\, S_n(1)|$, 
where $S_n(t) = \frac{1}{n} \sum_{i=1}^{\lfloor tn\rfloor} X_i$ is the partial sum process. 
If the noise terms are iid and the null hypothesis holds, then $\sqrt{n}T_n \convw \sigma\sup_{t\in[0,1]} |B(t)-tB(1)|$, where $\sigma^2=\var(\epsilon)$ and $B(t)$ is a standard Brownian motion.
Thus, quantiles of the limiting Kolmogorov distribution may serve as critical values. 

It has long been recognized that the assumption of iid noise is too restrictive for many sampling situations, e.g.\ for data arising in finance \citep{stuaricua2005}, climatology \citep{karl1995}, and power engineering \citep{weron2014,shao2017}. 
While the CUSUM statistic may still be used for nonstationary and dependent errors, the critical values need to be adjusted. 
In the presence of stationary autocorrelation, the limit is determined by the long-run variance $\sigma^2=\sum_{h=-\infty}^\infty \text{Cov}(\epsilon_0, \epsilon_h)$, which may be estimated by a HAC estimator \citep{newey1987,andrews1991}. 
See \citet{Aue2013} and \citet{horvath_change_2024} for an overview of changepoint testing under temporal dependence.
If the errors are furthermore nonstationary, the long-run variance changes over time, and \cite{zhou2013} suggested a multiplier bootstrap to adjust the critical values accordingly. 
This idea has found wide adoption, see e.g.\ \citet{Dette2018, bucher2020detecting,bonnerjee_gaussian_2024,kohne_at_2025,bai_inference_2026}.

Critical values based on either HAC estimators or block bootstrapping, both rely on the asymptotic consistency of the scale estimator, and thus neglect the estimation error of the latter. 
In view of the non-trivial choice of tuning parameters, especially the number of lags considered for long-run variance estimation, both bias and sampling variance of an HAC estimator can be substantial.
These estimation errors lead to unstable behavior in finite samples and inflated test sizes.
It has been suggested by \cite{horvath_ratio_2008} and \citet{Shao2010} to overcome this limitation by means of self-normalised statistics, which use an inconsistent variance estimator $\widehat{\sigma}^2_{\textsc{sn}}$; see also \cite{Shao2015}.
For a carefully selected estimator, $\sqrt{n}\widehat{\sigma}^2_{\textsc{sn}}\Rightarrow \sigma^2 \cdot V_{\textsc{sn}}(B)$, where the random term $V_{\textsc{sn}}(B)$ does not depend on $\sigma^2$. The standardized statistic converges under the null hypothesis, $T^{\textsc{CUSUM}}(S_n)/\widehat{\sigma}_{\textsc{sn}}(S_n) \Rightarrow T(B)/V_{\textsc{sn}}(B)$, and this pivotal limit explicitly accounts for the sampling variance of the denominator.

For nonstationary data, however, the multiplicative self-normalisation is insufficient because the distributional limit of the partial sum process is not a scaled Brownian motion.
This was pointed out by \cite{zhou2013} in the locally stationary framework of \citep{Dahlhaus1997, zhou2009, Dahlhaus2017}, and it also holds true for more general concepts of nonstationarity \citep{mies_sequential_2023,palm_central_2025}.
Here, we subsume these asymptotic frameworks by the following high-level assumption.

\begin{assumption}[Gaussian approximation]\label{ass:gaussian}
    There exist a standard Brownian motion $B$, and a sequence of continuous, strictly increasing functions $\Sigma_n:[0,1]\to[0,\infty)$, such that $\inf_n \Sigma_n(1)>0$, $\Sigma_n(1) / n \to 0$ and
    \begin{align*}
        \Delta_n := \sup_{t\in[0,1]} \left|\sqrt{n}[S_n(t)-m(t)]- B(\Sigma_n(t))\right| = \Oc_{\pr}(r_n \sqrt{\Sigma_n(1)}),
    \end{align*}
    where $m(t)=\int_0^t \mu(s)\, \diff s\approx \ex[S_n(t)]$, and $r_n\to 0$.
\end{assumption}

The function $\Sigma_n$ may be interpreted as a \emph{variance clock} of the partial sum process.
For stationary time series or iid data $\Sigma_n(t) = t\sigma^2$ is linear, while for locally stationary time series, $\Sigma_n(t)=\Sigma(t)=\int_0^t\sigma^2(s)\,\diff s$ does not depend on $n$.
In the latter case, $\sigma^2(s)$ is the local long-run variance, that is, the local long-run variance of $X_{i, n}$ is approximately $\sigma^2(i/n)$. 
The setup in Assumption \ref{ass:gaussian} allows more general forms of non-stationarity, for instance, local long-run variances of the form $\sigma^2(i/\sqrt{n})$; 
if this $\sigma^2(x) = o(x)$, as $x\to\infty$, then $\Sigma_n(1) / n \to 0$.
For detailed sufficient conditions, we refer to \citet{zhou2013, Karmakar2020, mies_strong_2025, bonnerjee_gaussian_2024}. 
Note that any functional central limit theorem also establishes this assumption, though without explicit rate $r_n$.

For the canonical case of a nonlinear variance clock $\Sigma_n$, the self-normalised CUSUM statistic fails to be pivotal. Existing alternatives combine local and global information to construct self-normalisers under local stationarity \citep{heinrichs2021,heinrichs2025}, whereas we propose an entirely global approach. This leads to a simpler construction of the normaliser and accommodates a broader range of nonstationary behavior. In the sequel, we present a new test statistic to allow for self-normalisation even under nonstationarity.

Two ideas are central for our approach: 
\begin{itemize}
    \item[(i)] We consider self-normaliser functionals $D$ which are invariant under \emph{clock-changes}. That is, $D(G\circ g) = D(G)$ for any stochastic process $G\in C[0,1]$, and any continuous, strictly increasing clock-change $g:[0,1]\to[0,1]$.
    \item[(ii)] 
    For such a functional $D$, the distribution of $D(B\circ\Sigma)$ only depends on the total variance $\Sigma(1)$, which is a scalar quantity. Thus,  self-normalisation yields a pivotal limit.
\end{itemize}

Section \ref{sec:method} presents the test statistic in detail, along an extension to multiple change points, and an investigation of its asymptotic consistency. 
Section \ref{sec:empirical} illustrates the finite sample performance in comparison to existing tests for nonstationary time series.

\section{Main Results}\label{sec:method}

\subsection{Self-normalised test statistic}
For $p \ge 0$, we consider the null hypothesis
\begin{equation*}
    \mathbb{H}_0^{(p)}:\mu~\text{has at most}~p~\text{jumps}
    \quad\text{against}\quad
    \mathbb{H}_1^{(p)}:\mu~\text{has more than}~p~\text{jumps}.
\end{equation*}
The case $p=0$ is the test of a constant mean against the presence of at least one change point, presented in the introduction.
Denote the integrated mean function as $m(t)=\int_0^t\mu(s)\, ds$.
If $\mu$ is piecewise constant with $p$ jumps, then $m$ is piecewise affine with $p+1$ affine pieces. 
This observation is the basis of the proposed tests.
To this end, introduce for any $r\in\N$ the class $\Gc_r$ of functions $g:[0,1]\to\R$ for which there exists a partition $0=\theta_0<\theta_1<\cdots<\theta_r=1$ such that $g(t) = a_j t + b_j$ on $(\theta_{j-1}, \theta_j]$, $j=1,\dots,r$. 
Note that this does not require continuity at the partition points. 
For $r \ge 0$, and $f:[0,1]\to\R$, define
\begin{equation*}
    N_r(f) = \inf_{g\in\Gc_{r+1}}\|f - g\|_\infty.
\end{equation*}
Since the cumulative mean $m$ is piecewise affine under the null hypothesis $\mathbb{H}_0^{(p)}$, we clearly have 
\begin{equation} \label{eqn:num}
    N_p(S_n)\;\leq\; \| S_n - m \|_\infty \; \approx\; \tfrac{1}{\sqrt{n}}\|B\circ \Sigma_n \|_\infty
     \;\overset{d}{=}\; \sqrt{\tfrac{\Sigma_n(1)}{n}} \|B\|_\infty.
\end{equation}
and the approximation error is controlled by Assumption \ref{ass:gaussian}.
The equality in distribution holds because the sup-norm is invariant under clock-changes, and Brownian motion is self-similar.
Thus, $N_p(S_n)$ tends to zero under the null hypothesis, by the assumption $\Sigma_n(1) / n \to 0$.
On the other hand, under the alternative $\mathbb{H}_1^{(p)}$, the mean $\mu$ has more than $p$ jumps, and $N_p(S_n)\approx N_p(m) > 0$. 
This shows that $N_p$ is a sensible test statistic for $\mathbb{H}_0^{(p)}$, and its distribution depends on a single unknown quantity $\Sigma_n(1)$.

To obtain a feasible test, instead of estimating $\Sigma_n(1)$, we construct a self-normaliser which is also invariant under clock-changes.
We introduce the larger reference class $\Fc_r$, for $r\in\N$, as the class of functions that are monotone on each element of a partition of $[0,1]$ into $r$ intervals.
Since every affine function is monotone on each of its affine pieces, $\Fc_r$ is a convenient enlargement of the piecewise affine class, but unlike affine functions, $\Fc_r$ is invariant under clock-changes.
For any $r\in\N$, and any $f$ as above, define
\begin{align*}
    D_r(f) = \inf_{g\in\Fc_{r+1}}\|f - g\|_\infty.
\end{align*}
This functional is time-change-invariant and homogeneous.
Moreover, if the mean function $\mu$ has at most $p$ change points, it serves as a lower bound to the affine fit as
\begin{align}
    N_{p+k}(S_n) \;\geq\; D_{2p+k}(S_n-m) \;\approx\; \tfrac{1}{\sqrt{n}} D_{2p+k}(B\circ \Sigma_n) \; \overset{d}{=}\; \sqrt{\tfrac{\Sigma_n(1)}{n}} D_{2p+k}(B). \label{eqn:denom}
\end{align}
The use of $\Fc_{k+2p+1}$ is due to a counting argument. 
Under $\mathbb{H}_0^{(p)}$, the integrated mean $m$ has at most $p+1$ affine pieces. 
Subtracting such a function from a $(p+k+1)$-piece affine approximation produces a piecewise affine function on the union of the two partitions, which has at most $k+2p+1$ pieces. 
Since each affine piece is monotone, the stochastic denominator can be bounded below by an approximation error over $\Fc_{2p+k}$.

Notably, \eqref{eqn:num} and \eqref{eqn:denom} reveal the same scaling factor, and this property may be used to design a self-normalised test statistics as follows. 
Let $K_n$ be a positive integer sequence and let $w_1,w_2,\ldots$ be non-negative weights. 
Define
\begin{equation}
    T_{p, n} = \max_{1\leq k\leq K_n} w_k \frac{N_p(S_n)}{N_{p+k}(S_n)}. \label{eqn:Tstat}
\end{equation}
The numerator measures lack of fit under a model with at most $p$ jumps, while the denominator measures lack of fit after allowing $k$ additional jumps. 
Large values of $T_{p,n}$ indicate that the fit of the $p$-jump model is substantially improved by allowing further change points.

Based on our derivations above, we use as reference the pivotal statistic
\begin{equation*}
    Q_p = \sup_{k\in\N} w_k \frac{\|B\|_\infty}{D_{2p+k}(B)},
\end{equation*}
where $B$ is a standard Brownian motion. 
Let $q_{1-\alpha, p}$ denote the $(1-\alpha)$-quantile of $Q_p$. Then, we reject $\mathbb{H}_0^{(p)}$ whenever
\begin{equation}  \label{eq:decision_rule}
    T_{p, n} > q_{1-\alpha, p}.
\end{equation}

The classes $\Gc_r$ and $\Fc_r$ have distinct roles. 
The class $\Gc_r$ is the model class and used to fit a model with $r$ constant segments. 
The class $\Fc_r$ is only used for calibration of the critical values which are valid uniformly for all possible variance clocks $\Sigma_n$.
Unlike piecewise affine functions, piecewise monotone functions are stable under clock-changes. More precisely, if $f$ is monotone and $g$ is a continuous strictly increasing bijection of $[0, 1]$, then $f \circ g^{-1}$ is monotone. Hence, $\Fc_r$ yields a time-change-invariant, and therefore pivotal, lower bound for the stochastic denominator
However, it is not possible to use $D_{p+k}$ as denominator in the test statistic \eqref{eqn:Tstat}, because there is no $k$ such that $D_{p}(B+m) \geq D_{p+k}(B)$, even if $m$ is piecewise affine. 
In particular, for $f\in\Fc_r$, the mapping $u\mapsto f(u)+u$ can have arbitrarily many monotone segments.

\subsection{Asymptotic guarantees} \label{sec:asymptotics}

To ensure asymptotic size control for $K_n\to\infty$, we need uniform control of the approximation errors over $k$. 
Available Gaussian approximation results for (piecewise) locally stationary processes yield rates such as $r_n=n^{-1/4}\log^2 n$ under suitable dependence and moment conditions; see, for example, \citet[Proposition 5]{zhou2013}.
A second key fact is that a Brownian motion can be approximated by $k$ monotone pieces with error of order $k^{-1/2}$.

\begin{proposition} \label{prop:bm_approximation}
    For each fixed $p \ge 0$, there exist almost surely finite random constants $0 < C_{1,p} \le C_{2,p} < \infty$ such that, for all sufficiently large $k$,
    \begin{equation*}
        C_{1,p}k^{-1/2} \le D_{2p+k}(B) \le C_{2,p}k^{-1/2}.
    \end{equation*}
\end{proposition}

As a consequence of this proposition, the unweighted ratio $\|B\|_\infty/D_{2p+k}(B)$ grows like $k^{1/2}$.
Hence, the weights in $T_{p, n}$ must be selected carefully as follows.

\begin{assumption}[Weights] \label{ass:tuning}
    The truncation sequence satisfies $K_n \to \infty$ and $r_n K_n^{1/2} \to 0$. The weights are non-negative and satisfy $k^{1/2}w_k \to 0$.
\end{assumption}

The condition $r_n K_n^{1/2} \to 0$ balances the quality of the Gaussian approximation against the number of models included in the maximum over $k\leq K_n$. 
The condition $k^{1/2}w_k \to 0$ prevents the statistic from being driven by arbitrarily fine segmentations. 
A typical choice is $w_k=k^{-1/2-\eta}$ with small $\eta>0$.
With this weighting scheme, we can control the type I and type II error of the test procedure.

\begin{theorem} \label{thm:main}
    Suppose that Assumptions \ref{ass:gaussian} and \ref{ass:tuning} hold. Then, under $\mathbb{H}_0^{(p)}$,
    \begin{equation*}
        \limsup_{n\to\infty} \pr(T_{p, n} > q_{1-\alpha, p}) \leq \alpha.
    \end{equation*}
    Moreover, if $\mu$ has $q>p$ jumps and $K_n\geq q-p$ eventually, then $\pr(T_{p, n} > q_{1-\alpha, p}) \to 1$.
\end{theorem}

The test for at least one change point is obtained by taking $p=0$, with Brownian reference statistic $Q_0$.
The additional tuning parameters $w_k$ and $K_n$ are not needed to account for dependence nor nonstationarity, but for the possibility of multiple change points under the alternative. 
The common self-normalised CUSUM statistic of \citet{Shao2010} fails to be consistent against more than one change point, because the self-normaliser only accounts for a single change point.
However, it has already been pointed out by \citet{Shao2010} that consistency against multiple change points can be achieved by localizing the test statistic to shorter sub-intervals, and this idea has been further elaborated by \citet{zhang_unsupervised_2018,zhao_segmenting_2022,cheng_general_2024}.
All these methods require stationarity.

Theorem \ref{thm:main} shows that an increasing number of change points can be handled, if the weights $w_k$ are chosen carefully.
On the other hand, if one accepts an upper bound on the number of change points, we may set $K_n=K$ constant. 
In this case $r_n$ may vanish arbitrarily slow in Assumption \ref{ass:gaussian}.
Indeed, Theorem \ref{thm:main} remains valid with the smaller quantiles $q_{1-\alpha, p, K_n}$ of the truncated reference statistic
\begin{equation*}
    Q_{p,K_n} = \sup_{1\leq k\leq K_n} w_k \frac{\|B\|_\infty}{D_{2p+k}(B)},
\end{equation*}
Letting $K_n \to \infty$ yields the limit $Q_p = \sup_{k \ge 1} w_k \|B\|_\infty / D_{2p+k}(B)$, which is almost surely finite whenever $w_k k^{1/2} \to 0$. 
Since $Q_{p,K}$ is pivotal, its quantiles can be tabulated, see Table \ref{tab:quantiles}.
For weights $w_k = \log(1.5+k) k^{-1/2-\eta}$, the convergence as $K\to\infty$ is rather slow, i.e.\ the quantiles of $Q_p$ are far larger than those of $Q_{p,K}$ for any practically relevant $K$.
Thus, the latter are preferable.

\begin{table}
    \centering
    \caption{Quantiles $q_{1-\alpha,p,K}$ of $Q_{p,K}$, based on $N = 10^5$ simulated Brownian motions on a grid of $n = 2000$ points, for different weighting schemes.}
    \label{tab:quantiles}
    \footnotesize
    \begin{tabular}{l|ccc|ccc|ccc}
        \toprule
        & \multicolumn{3}{|c|}{$p = 0$} & \multicolumn{3}{|c|}{$p = 1$} & \multicolumn{3}{|c}{$p = 2$} \\
        & $K=10$ & $K=20$ & $K=30$ & $K=10$ & $K=20$ & $K=30$ & $K=10$ & $K=20$ & $K=30$ \\ \midrule
        \addlinespace[.1cm]
        \multicolumn{10}{l}{\quad\textit{Panel A: $w_k = k^{-1/2-\eta}, \eta=0.001$}} \\ 
        \addlinespace[.1cm]
        $\alpha=0.1\%$ & 14.94 & 14.94 & 14.94 & 21.40 & 21.40 & 21.40 & 25.30 & 25.30 & 25.30 \\
        $\alpha=1\%$ & 10.96 & 10.99 & 11.02 & 15.74 & 15.74 & 15.74 & 19.54 & 19.54 & 19.54 \\
        $\alpha=5\%$ & 7.86 & 7.96 & 8.02 & 11.48 & 11.48 & 11.48 & 14.48 & 14.48 & 14.48 \\
        $\alpha=10\%$ & 6.53 & 6.67 & 6.79 & 9.63 & 9.63 & 9.64 & 12.25 & 12.25 & 12.25 \\
        \midrule
        \addlinespace[.1cm]
        \multicolumn{10}{l}{\quad\textit{Panel B: $w_k = \log(1.5+k)\,k^{-1/2-\eta}/\log(2.5), \eta=0.001$}} \\ 
        \addlinespace[.1cm]
        $\alpha=0.1\%$ & 29.17 & 41.83 & 49.93 & 31.89 & 43.07 & 50.59 & 34.82 & 44.00 & 52.58 \\
        $\alpha=1\%$ & 22.75 & 32.55 & 38.98 & 24.73 & 34.13 & 40.06 & 26.69 & 35.45 & 40.99 \\
        $\alpha=5\%$ & 17.21 & 25.54 & 30.95 & 18.82 & 26.61 & 31.70 & 20.49 & 27.39 & 32.54 \\
        $\alpha=10\%$ & 14.56 & 22.14 & 26.70 & 16.16 & 22.98 & 27.39 & 17.68 & 23.78 & 28.10 \\
        \bottomrule
    \end{tabular}
\end{table}

Theorem \ref{thm:main} also justifies the sequential procedure that tests $\mathbb{H}_0^{(0)},\mathbb{H}_0^{(1)},\dots,\mathbb{H}_0^{(p)}$ and stops at the first non-rejection. 
If the true number of jumps is $p^*\leq p$, then every underspecified null hypothesis $\mathbb{H}_0^{(q)}$, $q<p^*$, is rejected with probability tending to one. 
At the correctly specified null $\mathbb{H}_0^{(p^*)}$, the probability of further rejection is asymptotically bounded by $\alpha$, and the sequential procedure automatically accounts for the testing multiplicity.
Denoting by $\widehat{p}$ the first $p$ in this sequence such that $\mathbb{H}_0^{(p)}$ is rejected, we obtain a one-sided confidence region $[\widehat{p},\infty)$ for the number $p^*$ of change points.

\subsection{Computational Considerations}

Evaluation of the test statistic requires to compute
\begin{align*}
N_j(S_n)
& = \sqrt{n} \min_{0 = \theta_0 < \dots < \theta_{j} < \theta_{j+1} = 1} \max_{1\le i\le j+1} \min_{a_i, b_i \in \R} \sup_{t \in (\theta_{i-1}, \theta_i]} | S_n(t) - a_i t - b_i|.
\end{align*}
For a fixed partition, the inner minimizations over $(a_i,b_i)$ are independent across segments. The outer minimization over partitions can be implemented by dynamic programming, or by exhaustive search when the maximal number of segments is small. 

The critical values $q_{1-\alpha, p}$ and $q_{1-\alpha, p, K}$ are obtained by simulating standard Brownian motions and computing the corresponding errors of the piecewise monotone approximations, $D_{r}(B)$.
The following result allows an efficient calculation of the latter.

\begin{proposition}\label{prop:iso-range}
    Let $\Fc^\pm$ denote the sets of continuous, monotonously increasing and decreasing functions on $[0, 1]$, respectively, and $x$ a continuous function on $[0, 1]$. Then,
    \begin{equation*}
        \inf_{f\in \Fc^\pm} \sup_{t\in[0, 1]} |x(t) - f(t)| = \tfrac{1}{2} \sup_{0 \le s < t \le 1} \pm \{x(s) - x(t)\}.
    \end{equation*}
\end{proposition}
For $\Fc = \Fc^+ \cup \Fc^-$, a direct consequence of the proposition is
\begin{equation*}
    \inf_{f\in \Fc} \sup_{t\in[0, 1]} |x(t) - f(t)| = \tfrac{1}{2} \min\left\{\sup_{0 \le s < t \le 1} x(s) - x(t), \sup_{0 \le s < t \le 1} x(t) - x(s)\right\}.
\end{equation*}

\subsection{Continuous Alternatives}

The preceding theory is tailored to piecewise constant means, for which $m(t)=\int_0^t\mu(s)\,ds$ is piecewise affine. For smooth deviations from constancy, the denominator of the test statistic \eqref{eqn:Tstat} does not shrink, and the procedure is inconsistent, so other normalisations may be more natural.

One possibility is to estimate the integrated long-run variance by local block differences. For an even block length $h_n$ and $L_n=\lfloor n/h_n\rfloor$, a typical estimator is
\begin{equation*}
    \widehat{\Sigma} = \frac{1}{n} \sum_{i=0}^{L_n-1}
    \left(\sum_{t = i h_n + 1}^{i h_n + h_n / 2} X_{t, n} - \sum_{t=i h_n + h_n / 2 + 1}^{(i + 1) h_n} X_{t, n}
    \right)^2.
\end{equation*}
Under local stationarity, this estimates $\Sigma(1)$ even if $\mu$ varies smoothly \citep{kohne_at_2025}.
It can thus be used as normaliser such that under $\mathbb{H}_0^{(p)}$, we have $\sqrt{n}N_p(S_n)/\sqrt{\widehat{\Sigma}}\leq \sqrt{n}\|S_n-m\|_\infty/\sqrt{\widehat{\Sigma}}_\infty\approx \|B\|_\infty$, which is also pivotal.
This is, however, not a self-normalised statistic, and thus the estimation error of $\widehat{\Sigma}$ is neglected. 
Moreover, the tuning parameter $h_n$ needs to be chosen carefully.

Another possibility is the bivariate partial sum approach based on blocks of length $h_n$ \citep{heinrichs2025}. Conceptually, one constructs a two-parameter process $S_n(t,s)$, where $t$ is the argument of local partial sum processes over the blocks and $s$ retains the original timescale. In particular, $S_n(1, s) = S_n(s)$. Consider
\begin{equation*}
    N_0(S_n) = \sqrt{n} \inf_{c\in\R} \sup_{s\in[0,1]} |S_n(1, s) - s c|
\end{equation*}
and $H_n(t) = \sqrt{n}\{S_n(t,1)-tS_n(1,1)\}$. Under the null of a constant mean,
\begin{equation*}
    \frac{N_0(S_n)}{\sup_{t\in[0,1]}|H_n(t)|} \convw
    \frac{\inf_{c\in\R} \sup_{s\in[0,1]} |S(1, s) - s c|}{\sup_{t\in[0,1]} |S(t,1) - t S(1,1)|},
\end{equation*}
where $S$ is a Gaussian sheet limit. The denominator is a Brownian bridge functional, while the numerator measures deviation from linearity in the original time coordinate. Moreover, the two limiting components are independent. Replacing the numerator by its upper bound $\sup_s|S(1, s)|$ yields a pivotal conservative calibration involving an independent Brownian motion and Brownian bridge \citep[see][Corollary 8]{heinrichs2025}. This route targets general non-constant means rather than the number of jumps, and is therefore not pursued here.

\section{Empirical Results} \label{sec:empirical}

A Monte Carlo study was conducted to assess the finite sample performance of the proposed test.\footnote{Python code is available in the repository: \url{https://github.com/FlorianHeinrichs/PivotalSelfNormalization}} We generated data from $X_{i, n} = \mu(i/n) + \sigma(i/n) \eps_i$, for $i=1, \dots, n$, where $\mu$ is the mean function, $\sigma^2$ a (non-)constant variance and $(\eps_i)_{i\in\N}$ is a centred error process. We considered the null model $\mu_0(x)=0$, and the alternatives $\mu_1(x)=\id(x\ge1/2)$ and $\mu_2(x) = \id(x \ge 1/4) - \id(x\ge1/2) / 2 + \id(x \ge 3/4)$. For the variance, we used $\sigma_0(x) = 1/2, \sigma_1(x) = 1/4 + x/2$ and $\sigma_2(x) = x^{10}$. The first case corresponds to constant variance, whereas $\sigma_2$ is a stress-test design in which the variance mass is highly unevenly distributed over time. 

Two error processes were used. The first consists of independent standard normal errors. The second is locally stationary and given by
\begin{equation*}
    \eps_{i, n}^{(ls)} = \sqrt{a(i/n)} \eps_i^{(1)} + \sqrt{1 - a(i/n)} \eps_i^{(2)}, \qquad 
    a(t) = \tfrac{1}{2}\big\{1 - \cos\big(\tfrac{\pi}{2}\{-\cos(\pi t) + 1\}\big) \big\},
\end{equation*}
where $\eps_i^{(k)}=\sqrt{3/4} (\eta_i^{(k)} + \eps_{i-1}^{(k)}/2)$, $k=1,2$, with $\eta_i^{(1)}\sim \Nc(0,1)$ and $\eta_i^{(2)}\sim \Uc(-\sqrt3,\sqrt3)$.

For each configuration, 1000 replications were generated and the nominal level was $\alpha=5\%$. We compare the proposed test in \eqref{eq:decision_rule} with a test based on the block-difference estimator $\hat{\Sigma}$ (\texttt{LRV}), the bootstrap of \citet{zhou2013} (\texttt{Boot}), and self-normalised tests of \citet{Shao2010}, \citet{heinrichs2021} and \citet{heinrichs2025} (\texttt{SN1-3}). For the proposed test we used $K=10$ and weights $w(k) = \log(1.5 + k) k^{-1/2 - \eta}$ with $\eta = 0.001$. Accordingly, we used approximated quantiles of $Q_{p, K}$ rather than $Q_p$. Unlike the proposed test, the competing methods require bandwidth choices. We used the default choices recommended in the respective papers and $h_n=\lfloor n^{1/3}\rfloor$ for $\hat{\Sigma}$.

Empirical rejection rates (in \%, $\alpha = 5\%$) are reported in Table \ref{tab:results}. Under the null $\mu_0$, \texttt{SN2} and \texttt{SN3} largely hold the level, with \texttt{SN3} being extremely conservative. \texttt{SN1}, \texttt{Boot} and \texttt{LRV}, in contrast, overreject for locally stationary errors or strongly time-varying variance, and under $\sigma_2$ their size distortions persist as $n$ grows. The proposed test is conservative under $\sigma_0$ and $\sigma_1$. Under $\sigma_2$, it overrejects severely for $n=100$, but its size approaches the nominal level as $n$ increases, in line with its pivotal limit under non-stationarity.

Under $\mu_1$, \texttt{SN1}, \texttt{Boot} and \texttt{LRV} have power close to one, partly due to their size distortions. The proposed test is comparable for $n \ge 200$ and clearly more powerful than \texttt{SN2} and \texttt{SN3}. Multiple changes ($\mu_2$) separate the methods more clearly. \texttt{SN1} breaks down completely and \texttt{Boot} loses most of its power for $n \le 200$, while the power of the proposed test increases steadily with $n$. Apart from \texttt{LRV}, which controls size only for independent errors with $\sigma_0$ or $\sigma_1$, it is the most powerful method under $\mu_2$.

As discussed in Section \ref{sec:asymptotics}, the proposed test can sequentially test $\mathbb{H}_0^{(0)},\dots,\mathbb{H}_0^{(p)}$ and stop at the first non-rejection $\widehat{p}$, which is analysed empirically in the appendix.

{
\setlength{\tabcolsep}{2.5pt}
\begin{table}
\centering
\caption{Empirical rejection rates for various choices of $\mu$, $\sigma$ and $\eps$.}
\label{tab:results}
\footnotesize
\begin{tabular}{ll|rrrrrr|rrrrrr|rrrrrr}
	\toprule
	&& \multicolumn{6}{c}{$\mu = \mu_0$} & \multicolumn{6}{|c}{$\mu = \mu_1$} & \multicolumn{6}{|c}{$\mu = \mu_2$}\\
	$n$ & $\sigma$ & Ours & \texttt{SN1} & \texttt{SN2} & \texttt{SN3} & \texttt{Boot} & \texttt{LRV} & Ours & \texttt{SN1} & \texttt{SN2} & \texttt{SN3} & \texttt{Boot} & \texttt{LRV} & Ours & \texttt{SN1} & \texttt{SN2} & \texttt{SN3} & \texttt{Boot} & \texttt{LRV} \\
	\midrule
    \addlinespace[.1cm]
    \multicolumn{20}{l}{\quad\textit{Panel A: IID errors}} \\ 
    \addlinespace[.1cm]
	100 & $\sigma_0$ & 0.0 & 3.5 & 12.8 & 0.0 & 7.3 & 5.5 & 76.1 & 100.0 & 0.8 & 0.6 & 98.7 & 99.9 & 11.2 & 2.3 & 0.8 & 1.4 & 7.7 & 99.9 \\
    100 & $\sigma_1$ & 0.0 & 6.9 & 15.4 & 0.0 & 9.5 & 7.8 & 71.9 & 99.9 & 0.3 & 0.7 & 98.1 & 99.9 & 10.9 & 3.7 & 0.0 & 0.3 & 7.0 & 99.4 \\
    100 & $\sigma_2$ & 78.2 & 50.6 & 1.6 & 0.0 & 22.8 & 69.8 & 100.0 & 100.0 & 8.9 & 0.0 & 100.0 & 100.0 & 100.0 & 0.1 & 5.1 & 0.0 & 0.2 & 100.0 \\
    200 & $\sigma_0$ & 0.0 & 5.3 & 1.4 & 0.0 & 5.2 & 7.4 & 99.5 & 100.0 & 33.6 & 8.0 & 100.0 & 100.0 & 48.0 & 0.1 & 28.1 & 26.6 & 11.5 & 100.0 \\
    200 & $\sigma_1$ & 0.0 & 6.7 & 3.0 & 0.0 & 5.9 & 3.2 & 98.9 & 100.0 & 18.8 & 16.1 & 99.9 & 100.0 & 42.4 & 0.3 & 20.4 & 30.4 & 12.9 & 100.0 \\
    200 & $\sigma_2$ & 6.1 & 53.4 & 4.2 & 0.0 & 24.1 & 74.4 & 100.0 & 100.0 & 16.5 & 0.2 & 100.0 & 100.0 & 100.0 & 0.0 & 11.6 & 56.2 & 2.8 & 100.0 \\
    500 & $\sigma_0$ & 0.0 & 5.7 & 6.8 & 0.0 & 4.7 & 5.2 & 100.0 & 100.0 & 90.4 & 48.1 & 100.0 & 100.0 & 100.0 & 0.0 & 81.4 & 62.2 & 62.4 & 100.0 \\
    500 & $\sigma_1$ & 0.0 & 8.6 & 3.2 & 0.0 & 6.3 & 4.0 & 100.0 & 100.0 & 83.0 & 57.9 & 100.0 & 100.0 & 99.9 & 0.0 & 71.0 & 59.7 & 60.0 & 100.0 \\
    500 & $\sigma_2$ & 0.0 & 50.4 & 4.9 & 0.1 & 20.6 & 87.2 & 100.0 & 100.0 & 34.1 & 95.6 & 100.0 & 100.0 & 100.0 & 0.0 & 19.9 & 99.2 & 77.0 & 100.0 \\
	\midrule
    \addlinespace[.1cm]
    \multicolumn{20}{l}{\quad\textit{Panel B: Locally stationary errors}} \\ 
    \addlinespace[.1cm]
    100 & $\sigma_0$ & 0.0 & 7.8 & 6.3 & 0.0 & 8.6 & 34.9 & 83.7 & 100.0 & 1.3 & 0.5 & 94.7 & 100.0 & 25.6 & 5.8 & 1.1 & 1.1 & 6.4 & 100.0 \\
    100 & $\sigma_1$ & 0.1 & 14.0 & 0.7 & 0.0 & 12.4 & 39.0 & 84.5 & 98.0 & 0.3 & 0.3 & 91.4 & 100.0 & 40.1 & 8.7 & 0.1 & 1.0 & 7.5 & 100.0 \\
    100 & $\sigma_2$ & 95.2 & 61.0 & 1.1 & 0.0 & 28.0 & 81.4 & 100.0 & 99.8 & 7.6 & 0.0 & 100.0 & 100.0 & 100.0 & 2.4 & 1.9 & 0.0 & 2.1 & 100.0 \\
    200 & $\sigma_0$ & 0.0 & 8.1 & 2.2 & 0.0 & 7.5 & 27.4 & 99.1 & 100.0 & 22.2 & 12.2 & 99.7 & 100.0 & 54.7 & 1.3 & 19.6 & 19.0 & 13.7 & 100.0 \\
    200 & $\sigma_1$ & 0.0 & 12.9 & 0.0 & 0.0 & 8.2 & 25.6 & 96.2 & 99.8 & 7.9 & 7.6 & 96.6 & 100.0 & 50.9 & 3.5 & 5.6 & 21.1 & 13.4 & 100.0 \\
    200 & $\sigma_2$ & 33.5 & 55.6 & 1.9 & 0.0 & 29.4 & 80.2 & 100.0 & 99.9 & 6.7 & 0.3 & 100.0 & 100.0 & 100.0 & 0.2 & 3.2 & 63.1 & 10.8 & 100.0 \\
    500 & $\sigma_0$ & 0.0 & 9.2 & 4.2 & 0.0 & 6.1 & 17.2 & 100.0 & 100.0 & 58.6 & 67.8 & 100.0 & 100.0 & 99.1 & 0.1 & 45.9 & 68.8 & 39.3 & 100.0 \\
    500 & $\sigma_1$ & 0.0 & 15.1 & 1.5 & 0.3 & 7.6 & 12.1 & 99.9 & 100.0 & 16.0 & 70.9 & 100.0 & 100.0 & 89.3 & 0.5 & 10.4 & 74.9 & 30.9 & 100.0 \\
    500 & $\sigma_2$ & 2.4 & 53.9 & 3.1 & 0.8 & 23.9 & 82.6 & 100.0 & 100.0 & 3.4 & 93.4 & 100.0 & 100.0 & 100.0 & 0.0 & 1.7 & 95.6 & 65.9 & 100.0 \\
\bottomrule
\end{tabular}
\end{table}
}

\section*{Declaration of the use of generative AI and AI-assisted technologies}
During the preparation of this work the authors used Claude Opus 5.5 in order to refine the wording, double-check mathematical proofs, and identify additional relevant literature, in particular the use $h$-extrema for the bound in Proposition \ref{prop:bm_approximation} was suggested. After using this tool the authors reviewed and edited the content as necessary and take full responsibility for the content of the publication.

\subsection*{Acknowledgement}
FM acknowledges funding in the project VI.Veni.242.365, which is financed by the Dutch Research Council (NWO) under the grant  https://doi.org/10.61686/AGAFX90293. 

\appendix
\section{Proofs}

\begin{proof}[of Proposition \ref{prop:bm_approximation}]
    By Proposition 3.7 of \cite{picard2008}, for almost every path of the Brownian motion $B$ and $\eps > 0$, an approximation $B_\eps$ exists with $\| B - B_\eps\| \le \eps$ and $N_\eps \sim C \eps^{-2}$ monotone segments. In particular, $\inf_{f\in \Fc_{k+2p}} \|B - f\|_\infty \le C k^{-1/2}$. 
    
    In addition, we receive a lower bound from the study of $h$-extrema of Brownian motions \citep[see, e.\,g.,][]{schertzer2018}. More specifically, for $h > 0$ and a Brownian motion $B$, define the sequences
    \begin{align*}
        s_{2n+1} & = \inf\{ t \ge s_{2n} : B(t) - \inf_{s\in[s_{2n}, t]} B(s) = h\}, \\
        s_{2n+2} & = \inf\{ t \ge s_{2n+1} : \sup_{s\in[s_{2n+1}, t]} B(s) - B(t) = h\}, 
    \end{align*}
    and
    \begin{align*}
        t_{2n+1} & = \sup\{ t \in[s_{2n}, s_{2n+1}] : \inf_{s \in[s_{2n}, s_{2n+1}]} B(s) = B(t)\}, \\
        t_{2n+2} & = \sup\{ t \in[s_{2n+1}, s_{2n+2}] : \sup_{s\in[s_{2n+1}, s_{2n+2}]} B(s) = B(t)\}, 
    \end{align*}
    where $s_0, t_0 = 0$. Intuitively, $s_n$ delimits intervals in which $B$ increases (or decreases) by $h$ from its minimum (or maximum) in $[s_{n-1}, s_n]$. The points $t_n$ are the starting points of this increase (or decrease), in the sense that they are the last time instants where the minimum (or maximum) in $[s_{n-1}, s_n]$ is assumed. Put differently, $t_{2n+1}$ and $s_{2n+1}$ delimit the shortest interval of an $h$-increase after the last $h$-decrease (indicated by $t_{2n}$ and $s_{2n}$). 

    The $h$-extrema connect to the functional $D_m(B)$ as follows.
    Suppose that $f$ has $m$ monotone segments, i.e.\ $f\in \Fc_m$, and satisfies $\|B-f\|_\infty \leq \frac{h}{3}$.
    Suppose moreover that the path of $B$, on the interval $[0,1]$, admits $M(h)$ of the $h$-extrema as defined above, i.e.\ $M(h)=\sup\{k\,:\, s_k \leq 1\}$.
    By definition, $|B(t_i)-B(s_{i})| \ge h$ and $\mathrm{sign}(B(t_{i}) - B(s_i))$ alternates from one interval to the next. 
    Since $f$ can vary from $B$ at most $h/3$, $f(t_{i}) -f(s_i)$ has the same sign as $B(S_{i+1}) - B(S_i)$, so the monotonicity direction of $f$ alternates between intervals. 
    This shows that $m\geq M(h)$, or equivalently,
    \begin{align*}
        \{\inf_{f\in\Fc_m}\|B-f\|_\infty >h  \} \supseteq \{ M(3h)> m\}. 
    \end{align*}
    We may thus derive stochastic lower bounds on $D_m(B)$ by finding stochastic lower bounds for the random variables $M(h)$.
    Specifically, we use the inequality
    \begin{align*}
        P(D_m>h) \geq P(M(3h)>m) \geq P\left(\sum_{i=1}^m |s_{i+1}(3h)-s_i(3h)| \leq \tfrac{1}{2}  \right).
    \end{align*}
    
    It has been shown by \citet{neveu2006} that the random variables $\Delta_i(h)=s_{i+1}(h)-s_i(h)$ are iid with Laplace transform $\ex e^{-\lambda \Delta_i(h)}= 1/\cosh(h \sqrt{2\lambda})$, and expectation $\ex(\Delta_i(h))=h^2$.
    A standard calculation \citep[see e.\,g.][Exercise~7.5.1]{durrett_probability_2019}, shows that this corresponds to the distribution of the first hitting time of the Brownian motion,
    $\tau(h)=\inf\{t>0:|B(t)|=h\}$.
    More specifically, note that $1=\ex \exp(z B_{\tau(h)}-z^2 \tau(h)/2)$ by optional stopping of the exponential martingale. Substituting $z=\sqrt{2\lambda}$ yields the corresponding Laplace transform, since $B_{\tau(h)} = \pm h$ with equal probability.
    Now compute the distribution of $\tau(h)$ as
    \begin{align*}
        P(\tau(h)>v) = P\left(\sup_{s\in[0,v]} |B_s|<h\right) = P\left(\sup_{s\in[0,1]} |B_s|< h/\sqrt{v}\right) 
        &= \frac{4}{\pi} \sum_{k=1}^\infty \frac{(-1)^{k+1}}{2k-1}\exp\left(-\frac{(2k-1)^2 \pi^2}{8(h/\sqrt{v})^2}\right) \\
        &\leq  \frac{4}{\pi} \exp\left(-\frac{v}{h^2} \frac{\pi^2}{8}\right).
    \end{align*}
    The series formula is due to \citet{erdos_certain_1946}, Theorem 2 therein, and the last inequality holds for all alternating series of strictly decreasing sequences.
    
    We do not need the exact distribution of $\tau(h)$, but only the fact that $\tau(h)/h^2$ is sub-exponential, uniformly in $h$, and its distribution does not depend on $h$.
    We may then employ the Bernstein inequality for independent sub-exponential random variables \citep[Thm.~2.9.1]{vershynin_high-dimensional_2026} to the centered sum $\sum_{i=1}^m [\Delta_i(h)-h^2]$, which yields for a universal $C$,
    \begin{align*}
        P\left( \left|\sum_{i=1}^m [\Delta_i(h)-h^2]\right| > h^2 v  \right) \leq 2 \exp\left(-C \frac{v^2}{m+ v}\right).
    \end{align*}
    Putting things together, and substituting $h=c/\sqrt{m}$ for some $c>0$ such that $\delta=\frac{1}{2}-9c^2>0$,
    \begin{align*}
        P(D_m\leq c/\sqrt{m}) 
        \leq P\left( \sum_{i=1}^m \Delta_i(\tfrac{3c}{\sqrt{m}}) > \tfrac{1}{2} \right)
        &= P\left( \sum_{i=1}^m \left[\Delta_i(\tfrac{3c}{\sqrt{m}}) - \tfrac{9c^2}{m}\right] > \tfrac{1}{2} - 9c^2 \right) \\
        &\leq \exp\left( - C \frac{\delta^2/(3h)^4}{m+\delta/(3h)^2} \right) \\
        &= \exp\left( -\tilde{C} \frac{\delta^2 m^2}{m+m\delta} \right) 
        \quad = \exp\left( -\tilde{C} \frac{\delta^2 m}{1+\delta} \right).
    \end{align*}
    The latter is summable in $m$, showing that $\limsup_{m\to\infty} \sqrt{m}D_m > c$ almost surely by the Borel-Cantelli lemma.
\end{proof}

The next lemma is the uniform continuous mapping step needed to treat the statistic with growing $K_n\to\infty$.

\begin{lemma} \label{lem:conv_statistic}
    Let $x_n$ be random elements of $\ell^\infty([0,1])$ such that $\|x_n - G_n\|_\infty =\Oc_{\pr}(\rho_n)$, where $\rho_n K_n^{1/2}\to0$, $G_n(t)=B(\Sigma_n(t))$, and $\Sigma_n$ is continuous strictly increasing with $\Sigma_n(0)=0$.
    If $k^{1/2} w_k \to 0$, then, for each fixed $p \ge 0$,
    \begin{equation*}
        \max_{1\leq k\leq K_n} w_k \frac{\|x_n\|_\infty}{D_{p+k}(x_n)} \convw Q_p.
    \end{equation*}
\end{lemma}

The proof uses the Lipschitz property of $x\mapsto D_{p+k}(x)$ with respect to the sup-norm, the lower bound in Proposition \ref{prop:bm_approximation}, and the Brownian time-change identity

\begin{equation} \label{eq:time_change}
    \frac{\|B\circ \Sigma_n\|_\infty}{D_{p+k}(B\circ\Sigma_n)} \stackrel{\Dc}{=} \frac{\|B\|_\infty}{D_{p+k}(B)}.    
\end{equation}
Indeed, both numerator and denominator are multiplied by the same factor $\Sigma_n(1)^{1/2}$ after the time change.

\begin{proof}[of Lemma \ref{lem:conv_statistic}]
	By a time-shift and self-similarity of the Brownian motion,
	\begin{equation} \label{eq:factorization_num}
		\|G_n\|_\infty
		= \sup_{t\in [0, 1]} \big| B\big(\Sigma_n(t)\big)\big|
		= \sup_{t\in [0, \Sigma_n(1)]} |B(t)| 
		= \sup_{t\in [0, 1]} |B(\Sigma_n(1) t)|
		\stackrel{\Dc}{=} \sqrt{\Sigma_n(1)} \|B\|_\infty,
	\end{equation}
	In the following, we show by the same self-similarity argument and time-shift,
	\begin{equation} \label{eq:factorization_denom}
		\inf_{f \in \Fc_k} \| G_n - f\|_\infty \stackrel{\Dc}{=} \sqrt{\Sigma_n(1)} \inf_{f \in \Fc_k} \| B - f\|_\infty.
	\end{equation}
	By assumption, $\Sigma_n(t)$ is bijective with inverse $\Sigma_n^{-1}(t)$. Since $G_n(t) = B(\Sigma_n(t))$, we may rewrite
	\begin{align*}
		\inf_{f \in \Fc_k} \| G_n - f\|_\infty
		& = \inf_{0 = s_0 < \dots < s_k= 1}  \inf_{f_1, \dots, f_k \in \Fc_1} \max_{i=1}^{k} \sup_{t\in[s_{i-1}, s_i]}|G_n(t) - f_i(t)| \\
		& = \inf_{0 = s_0 < \dots < s_k= 1}  \inf_{f_1, \dots, f_k \in \Fc_1} \max_{i=1}^{k} \sup_{t\in[\Sigma_n(s_{i-1}), \Sigma_n(s_i)]}|B(t) - f_i(\Sigma_n^{-1}(t))| \\
		& \stackrel{\Dc}{=} \sqrt{\Sigma_n(1)} \inf_{0 = s_0 < \dots < s_k= 1}  \inf_{f_1, \dots, f_k \in \Fc_1} \max_{i=1}^{k} \sup_{t\in[\Sigma_n(s_{i-1})/\Sigma_n(1), \Sigma_n(s_i)/\Sigma_n(1)]}|B(t) - f_i(t)| \\
		& = \sqrt{\Sigma_n(1)} \inf_{0 = t_0 < \dots < t_k= 1}  \inf_{f_1, \dots, f_k \in \Fc_1} \max_{i=1}^{k} \sup_{t\in[t_{i-1}, t_i]}|B(t) - f_i(t)| \\
		& = \sqrt{\Sigma_n(1)} \inf_{f \in \Fc_k} \| B - f\|_\infty,
	\end{align*}
	where the second equality is due to a time shift, the third equality (in distribution) due to self-similarity of the Brownian motion and monotonicity of $f_i(\Sigma_n^{-1}(\Sigma_n(1) t))$, and the fourth equality due to the substitution $t = \Sigma_n(s)/\Sigma_n(1)$. Note that the arguments do not depend on $k$, and indeed can be applied simultaneously for $k\in\N$, such that \ref{eq:factorization_denom} holds uniformly in $k$.
	
	Fix some $p \ge 0$ and let $\nu_n = \Sigma_n(1)^{1/2}$, which is bounded from below by Assumption \ref{ass:gaussian}. 
    By \eqref{eq:factorization_denom} and Proposition \ref{prop:bm_approximation}, 
	\begin{align*}
		D_* := \nu_n^{-1} \inf_{k\ge 1} \sqrt{k} \inf_{f\in\Fc_k} \| G_n - f \|_\infty \overset{\Dc}{=} \inf_{k\ge 1} \sqrt{k} \inf_{f\in\Fc_k} \| B - f \|_\infty > 0 \quad \text{almost surely.}
	\end{align*}
	Let $\Delta_n = \| x_n - G_n\|_\infty / \nu_n$. For $x \in \ell^\infty([0, 1])$, define
	\begin{equation*}
		T_K(x):=\max_{1\le k\le K} w_k \frac{\|x\|_\infty}{D_{p+ k}(x)}, \qquad T(x):= \sup_{k\in \N} w_k \frac{\|x\|_\infty}{D_{p+ k}(x)}.
	\end{equation*}
	The functional $D_{p+k}$ is Lipschitz with respect to the sup norm. 
	For $c>0$ and $M<\infty$, define the event 
	\begin{equation*}
		A_n(c,M) = \left\{ D_* \ge c,\; \|G_n\|_\infty \le M \nu_n ,\;  \Delta_n\sqrt{K_n}\le c / 2 \right\}.
	\end{equation*}
	By assumption, the probability of the complement of $A_n(c,M)$ can be made arbitrarily small. More specifically, for any $\eta > 0$, we can select $c > 0$ and $M < \infty$, such that $\limsup_{n\to\infty} 1 - \pr(A_n(c, M)) \le \eta$.
	
	On the event $A_n(c, M)$, $D_{p+k}(G_n) \ge c k^{-1/2} \nu_n$, for $1\le k\le K_n$. 
	By Lipschitz continuity, $D_{p+ k}(x_n) \ge D_{p+ k}(G_n) - \nu_n \Delta_n$.
	Since $k \le K_n$, 
	\begin{equation*}
		\Delta_n \le \frac{c}{2\sqrt{K_n}} \le \frac{c}{2\sqrt{k}},
	\end{equation*}
	and therefore
	\begin{equation} \label{eq:lower_bound_D}
		D_{p+ k}(x_n)\ge \nu_n \bigg(\frac{c}{\sqrt{k}}-\frac{c}{2\sqrt{k}}\bigg) = \nu_n \frac{c}{2\sqrt{k}}.
	\end{equation}
	Hence on $A_n(c,M)$, all denominators appearing in $T_{K_n}(x_n)$ are strictly positive.
	Now define $V_k(x)=w_k \|x\|_\infty / D_{p+ k}(x)$, such that $T_K(x)=\max_{1\le k\le K} V_k(x)$. For $k\le K_n$, on $A_n(c,M)$,
	\begin{align*}
		|V_k(x_n)-V_k(G_n)| &= w_k\bigg| \frac{\|x_n\|_\infty}{D_{p+ k}(x_n)} - \frac{\|G_n\|_\infty}{D_{p+ k}(G_n)} \bigg| \\
		&\le w_k \frac{\big|\|x_n\|_\infty-\|G_n\|_\infty\big|}{D_{p+ k}(x_n)} + w_k \|G_n\|_\infty \Big| \frac{1}{D_{p+ k}(x_n)}-\frac{1}{D_{p+ k}(G_n)} \Big|.
	\end{align*}
	By \eqref{eq:lower_bound_D} and the corresponding bound on $D_{p+k}(G_n)$, 
	\begin{equation*}
		\Big| \frac{1}{D_{p+ k}(x_n)}-\frac{1}{D_{p+ k}(G_n)} \Big|
		= \frac{|D_{p+ k}(x_n) - D_{p+ k}(G_n)|}{D_{p+ k}(x_n) D_{p+ k}(G_n)}
		\le \frac{2k}{c^2 \nu_n^2} |D_{p+ k}(x_n) - D_{p+ k}(G_n)|.
	\end{equation*}	
	Using $\big|\|x_n\|_\infty-\|G_n\|_\infty\big| \le \Delta_n \nu_n$, and $|D_{p+ k}(x_n)-D_{p+ k}(G_n)|\le\Delta_n \nu_n$, we obtain 
	\begin{align*}
		|V_k(x_n)-V_k(G_n)| \le w_k\frac{\Delta_n}{c/(2\sqrt{k})} + w_k M \frac{2k \Delta_n}{c^2}  
		= \frac{2w_k\sqrt{k}}{c}\Delta_n + \frac{2M w_k k}{c^2}\Delta_n.
	\end{align*}
	Since $w_k\le C_w k^{-1/2}$, 
	\begin{equation*}
		|V_k(x_n)-V_k(G_n)| \le \frac{2C_w}{c}\Delta_n + \frac{2M C_w}{c^2}\Delta_n\sqrt{k} \le \frac{2C_w}{c}\Delta_n + \frac{2M C_w}{c^2}\Delta_n\sqrt{K_n}.
	\end{equation*}
	By taking the maximum over $k$, 
	\begin{equation*}
		|T_{K_n}(x_n)-T_{K_n}(G_n)| \le \frac{2C_w}{c}\Delta_n + \frac{2M C_w}{c^2}\Delta_n\sqrt{K_n},
	\end{equation*}
	which converges to $0$ by assumption on $K_n$. 
	Now, since $D_*>0$ a.s., $D_{p+k}(G_n)\ge D_* k^{-1/2} \nu_n$, such that
	\begin{equation*}
		\frac{w_k}{D_{p, k}(G_n)} \le \frac{C_w k^{-1/2}}{D_* k^{-1/2} \nu_n} = \frac{C_w}{D_* \nu_n}.
	\end{equation*}
	Since $\|G_n\|_\infty / \nu_n<\infty$ a.s., $T(G_n)\le \|G_n\|_\infty C_w / (D_* \nu_n) <\infty$ a.s. 
	Moreover, by \eqref{eq:factorization_num} and \eqref{eq:factorization_denom}, 
		\begin{equation*}
		T_{K_n}(G_n) = \max_{1\le k\le K_n} w_k \frac{\|G_n\|_\infty}{D_{p+ k}(G_n)} \stackrel{\Dc}{=} \max_{1\le k\le K_n} w_k \frac{\|B\|_\infty}{D_{p+ k}(B)} \to \sup_{k\in\N} w_k \frac{\|B\|_\infty}{D_{p+ k}(B)} = Q_p,
	\end{equation*}
	almost surely as a monotone sequence, as $K_n\to\infty$. Hence,
	\begin{equation*}
		T_{K_n}(x_n) = \big[T_{K_n}(x_n)-T_{K_n}(G_n)\big] + T_{K_n}(G_n) \convw Q_p.
	\end{equation*}	
\end{proof}

\begin{proof}[Theorem \ref{thm:main}]
    Let $\mu$ be piecewise constant with at most $p$ jumps, then $m \in \Gc_{p+1}$. Moreover, if $m \in \Gc_{p+1}$ and $f \in \Gc_{k+1}$, then $\sqrt{n} m + f\in \Gc_{p + k + 1}$.
    Denote $\tilde{S}_n = \sqrt{n}(S_n-m)$
    On one hand, 
    \begin{equation*}
        N_p(S_n) \le \|\sqrt{n} (S_n - m)\|_\infty  = \| \tilde{S}_n \|_\infty,
    \end{equation*}
    while on the other hand,
    \begin{equation*}
        N_k(S_n) = \inf_{f \in \Gc_{k+1}}\! \sqrt{n} \| S_n - f\|_\infty
        = \inf_{f \in \Gc_{k+1}} \|\tilde{S}_n + \sqrt{n}m - f\|_\infty 
        \ge \inf_{f \in \Gc_{k+p+1}} \| \tilde{S}_n - f\|_\infty
        \ge \inf_{f \in \Fc_{k+p+1}} \| \tilde{S}_n - f\|_\infty,
    \end{equation*}
    for any $k\in \N$. In particular,
    \begin{align*}
        T_{p, n}
        = \max_{1\le k \le K_n} w_k \frac{N_p(S_n)}{N_{p+k}(S_n)}
        \le \max_{1\le k \le K_n} w_k \frac{\|\tilde{S}_n\|_\infty}{D_{2p+k}(\tilde{S}_n)} 
        \convw Q_p
    \end{align*}
    by Lemma \ref{lem:conv_statistic}, hence, $\limsup_{n\to\infty} \pr(T_{p, n} > q_{1-\alpha, p}) \le \alpha$. 

    In contrast, if $\mu$ is not piecewise constant with $p$ or less jumps, then $m\notin \Gc_{p+1}$, and
    \begin{equation*}
        N_p(S_n) 
        = \inf_{f\in \Gc_{p+1}}\|\sqrt{n}(S_n - m) + \sqrt{n}(m - f)\|_\infty
        \ge \sqrt{n} \inf_{f\in \Gc_{p+1}}\|m - f\|_\infty - \|\tilde{S}_n\|_\infty \to\infty.
    \end{equation*}
    Under the alternative $H_1^{(p)}$, some $k^*\in\N$ exists with $m \in \Gc_{p+k^*}$, such that 
    \begin{equation*}
        \frac{N_p(S_n)}{N_{k+p}(S_n)} \ge \frac{\sqrt{n}\inf_{f\in\Gc_{p+1}}\|m - f\|_\infty - \|\tilde{S}_n\|_\infty}{\|\tilde{S}_n\|_\infty} \to \infty,
    \end{equation*}
    for any $k\ge k^*$, proving consistency.
    Note that the latter divergence holds because $\|\tilde{S}_n\|_\infty/\sqrt{\Sigma_n(1)} \overset{\Dc}{=}  \sup_{u\in[0,1]}|B(u)| + \Oc_{\pr}(r_n)$ by Assumption \ref{ass:gaussian}, and $\Sigma_n(1)=o(n)$.
\end{proof}

\begin{proof}[of Proposition \ref{prop:iso-range}]
    Let $m = 1/2 \sup_{0 \le s < t \le 1} x(s) - x(t)$ and define $f^*(t) = \sup_{s \le t} x(s) - m$. By continuity of $x$, $f^* \in \Fc$. By definition, $f^*(t) \ge x(t) - m$, hence $x(t) - f^*(t) \le m$.

    By definition of $m$, $x(s) - x(t) \le 2m$, for any $s \le t$, hence, $x(s) - m \le x(t) + m$. Taking suprema over $s \in [0, t]$, $f^*(t) \le x(t) + m$, such that $\|x - f^*\|_\infty \le m$. Since $f^* \in \Fc$, $\inf_{f\in \Fc} \|x -f\|_\infty \le m$. 

    Contrarily, let $f\in\Fc$ be arbitrary. Then, for all $t \in [0, 1]$, $x(t) \le f(t) + \|x - f \|_\infty$ and $f(t) \le x(t) + \|x - f \|_\infty$. By monotonicity of $f$,
    \begin{equation*}
        x(s) \le f(s) + \|x - f \|_\infty \le f(t) + \|x - f \|_\infty \le x(t) + 2 \|x - f \|_\infty,
    \end{equation*}
    for any $s \le t$, so $1/2 (x(s) - x(t)) \le \|x - f \|_\infty$. Taking suprema over $s < t$ on both sides yields $m \le \|x - f \|_\infty$. The first part of the proposition follows by taking infima over $f \in \Fc$. The second part follows by analogous arguments.
\end{proof}

\section{Additional Empirical Results} \label{app:empirical}

As outlined in Section \ref{sec:asymptotics}, the proposed test can be used to sequentially test the hypotheses $\mathbb{H}_0^{(0)},\mathbb{H}_0^{(1)},\dots,\mathbb{H}_0^{(p)}$ and stop at the first non-rejection $\widehat{p}$. The interval $[\widehat{p}, \infty)$ then defines a one-sided confidence region with nominal coverage probability $1-\alpha$ for the true number of changes. We illustrate this for the three-change alternative $\mu_2$ with locally stationary errors, testing $\mathbb{H}_0^{(p)}$ for $p=0,1,2,3$ at level $\alpha=5\%$. Table \ref{tab:hist} shows the frequencies of $\widehat{p}$ over 1000 replications. The category $\widehat{p} \ge 4$ contains the replications in which all four hypotheses are rejected, that is, in which the confidence region misses the true number of changes.

For $\sigma_0$ and $\sigma_1$, the confidence region covers the true number of changes in all replications, but the lower bound $\widehat{p}$ is conservative. For $n \le 200$, $\widehat{p}$ is either 0 or 1, with $\widehat{p}=0$ in 45-74\% of the replications. Even for $n=500$, it equals 1 in 89-98\% of the replications. In larger samples, the procedure thus reliably detects the presence of a change, but rarely more than one. In the $\sigma_2$ design, coverage is poor for $n=100$ (17.3\%), which mirrors the severe overrejection of the test in this setting (cf.\ Table \ref{tab:results}). Coverage is close to the nominal level for $n=200$ (94.7\%) and complete for $n=500$. For $n \ge 200$, $\widehat{p}$ is concentrated on 2 and 3, so the confidence region is considerably more informative than under $\sigma_0$ and $\sigma_1$.


\begin{table}
    \centering
    \caption{Frequencies of $\widehat p$ in 1000 replications for $\mu = \mu_2$ with locally stationary errors.}
    \label{tab:hist}
    \footnotesize
    \begin{tabular}{l|rrrrr|rrrrr|rrrrr}
        \toprule
        & \multicolumn{5}{|c}{$\sigma = \sigma_0$} & \multicolumn{5}{|c}{$\sigma = \sigma_1$} & \multicolumn{5}{|c}{$\sigma = \sigma_2$} \\
        $n$ & 0 & 1 & 2 & 3 & $\ge$4 & 0 & 1 & 2 & 3 & $\ge$4 & 0 & 1 & 2 & 3 & $\ge$4 \\
        \midrule
        100 & 744 & 256 & 0 & 0 & 0 & 599 & 401 & 0 & 0 & 0 & 0 & 0 & 0 & 173 & 827 \\
        200 & 453 & 547 & 0 & 0 & 0 & 491 & 509 & 0 & 0 & 0 & 0 & 0 & 160 & 787 & 53 \\
        500 & 9 & 984 & 7 & 0 & 0 & 107 & 892 & 1 & 0 & 0 & 0 & 0 & 676 & 324 & 0 \\
        \bottomrule
    \end{tabular}
\end{table}


\bibliographystyle{biometrika}
\bibliography{bibliography}

\end{document}